\documentclass[10pt,a4paper]{article}

\usepackage{tikz}
\usetikzlibrary{decorations.pathmorphing,calc}
\usetikzlibrary{positioning}
\usepackage{jheppub_kim}
\usepackage{pdflscape}
\usepackage{amsmath}
\usepackage{amssymb}
\usepackage{dcolumn}
\usepackage{bm}
\usepackage{color}
\usepackage{amsmath}
\usepackage{breqn}
\usepackage{graphicx}
\usepackage{epsfig}
\usepackage{caption,subcaption}
\usepackage{amsfonts}
\usepackage{graphicx}
\usepackage{parskip}
\usepackage{dcolumn}
\UseRawInputEncoding

\begin{document}


\title{Study of Wormhole in $f(Q)$ gravity with some dark energy models}

\author[a]{Tanmoy Chowdhury,}
\affiliation[a]{Department of Mathematics, Jadavpur University, Kolkata 700032, West Bengal, India.}
\emailAdd{tanmoych.ju@gmail.com}

\author[b]{Prabir Rudra,}
\affiliation[b]{Department of Mathematics, Asutosh College,
Kolkata-700 026, India.}
\emailAdd{prudra.math@gmail.com}

\author[c]{Tandrima Chowdhury,}
\affiliation[c]{Department of Mathematics, Jadavpur University, Kolkata 700032, West Bengal, India.}
\emailAdd{tandrimachowdhury9678@gmail.com}

\author[d]{Farook Rahaman}
\affiliation[d]{Department of Mathematics, Jadavpur University, Kolkata 700032, West Bengal, India.}
\emailAdd{rahaman@associates.iucaa.in}


\abstract{This study discusses the development of some particular static wormhole models in the background of an extended $f(Q)$ gravity theory. Wormhole solutions are derived by considering the radial pressure to admit an equation of state corresponding to Chaplygin gas. The Chaplygin gas equation of state is taken into consideration in two different forms: $p_{r}=-\frac{Bb(r)^{u}}{\rho^{a}}$, $p_{r}=-\frac{B}{\rho^{a}}$. Wormhole models are also generated assuming that a variable barotropic fluid may explain the radial pressure given by $p_{r} =-\omega\rho b(r)^u$. For every model, the shape function $b(r)$ is the function that can be derived from the wormhole metric in any scenario. The stability analysis of the wormhole solutions and the shape function viability for each situation are then investigated.Since each wormhole model is shown to violate the null energy condition (NEC), it can be understood that these wormholes are traversable. More generally, we investigate whether the model is stable under the hydrostatic equilibrium state condition using the TOV equation.The physical characteristics of these models are shown under the same energy circumstances. The typical characteristic is the radial pressure $p_{r}$ near the wormhole throat, which violates the NEC $(\rho+P_{r} \geq0)$. In some models, it is possible to meet the NEC at the neck and yet violate the DEC $(\rho - P_{r}\geq0)$. In summary, precise wormhole models may be generated, provided that $(\rho\geq0)$, and there may be a potential breach of the NEC at the wormhole's throat.}

\maketitle

\section{Introduction}
The development of accurate wormhole (WH) models has long been a difficult challenge in General Relativity (GR) and in modified gravity theories. The study of wormhole structures represents a highly dynamic and evolving area of research. 
The idea of wormholes as theoretical structures that link two parallel universes or two distant points in space and time has been a topic of extensive research in theoretical physics. The word ``Wormhole" was initially introduced by John Archibald Wheeler \cite{wheeler1957} to refer to the particles of the spacetime quantum foam between various parts of spacetime at the Planck scale.  As clearly explained in \cite{fullerwheeler1962}, these primitive wormhole solutions were hardly traversable and disintegrated shortly after their proposal. In spite of this, recent interest in wormhole theories was sparked after following the seminal work of Morris and Thorne in 1988 \cite{morristhorne1988} where the traversable wormhole process was considered and addressed by the authors. In this work, they considered static and spherically symmetric line elements. Unless matter and radiation move openly in all directions and for an appropriate duration of time, a wormhole is not considered to be traversable. A wormhole, proposed by Morris, Thorne, and Yurtsever in \cite{m88}, is shown to evolve into a time machine, resulting in causality violation \cite{visser1995,lobo2007}.

In general relativity, it is widely understood that exotic matter whose stress-energy tensor contradicts the null energy condition (NEC) \cite{morristhorne1988,visser1995}, under the requirements of the geometrical structure, supports wormhole spacetimes. All of the averaged and point-wise energy criteria are altered by traversable wormholes \cite{lobo2007}. On the other hand, the authors of the latest work \cite{konoplya2021} recently discovered solutions that explain traversable wormholes that are asymmetric, asymptotically flat, and supported by regular Dirac and Maxwell fields. Several claims have been proposed with violation of energy conditions since the necessity of exotic matter is problematic in the development of wormhole physics. Several arguments involve studying scalar-tensor theories or invoking quantum fields in curved spacetime \cite{visser1995}. These attempts have been made to lower the use of exotic matter.

One of the most popular approaches in this direction is the "volume integral quantifier" \cite{visser2003,kar2004}, which quantifies the total amount of energy condition-violating matter required to maintain the wormhole structure. This method provides a useful framework for evaluating the feasibility of constructing traversable wormholes with reduced amounts of exotic matter. In recent years, modified theories of gravity, including $f(Q)$ gravity, have seen extensive research on wormhole solutions. The non-metricity scalar $Q$ is generalized as $f(Q)$ in the symmetric teleparallel equivalent of general relativity, replacing the Ricci scalar $R$ used in Einstein's general relativity \cite{p17}. The choice of matter content threading the wormhole in the investigation of wormhole solutions in $f(Q)$ gravity is an important aspect. Two commonly considered matter models are the Chaplygin gas and barotropic gas. The Chaplygin gas is an exotic fluid with an equation of state that exhibits a transition from a positive pressure (dust-like) regime to a negative pressure (dark energy-like) regime \cite{f17}. The barotropic gas is a more general fluid with a linear equation of state relating the pressure and energy density \cite{n19}.

According to this perspective, various methods have been put out to address the issue. But we are unable to directly force the aforementioned requirement on the situation. The exotic form of matter or modified gravity theories, where the features of wormholes are provided by higher-order curvature factors, can be examined to address this issue. The investigation of traversable and thin-shell wormholes with their properties in $f(R)$ gravity has been conducted in this context, as documented by \cite{lobo2009}. In \cite{moraes2017}, Various assumptions have been made on the radial and lateral pressure relations when exploring the wormhole geometries in $f(R, T)$ gravity. The form function solutions were also discovered and their characteristics concerning energy conditions were discussed. The WH solutions were examined in the context of $R^2$-gravity and exponential $f(R, T)$ formalism by Moraes and his associates \cite{moraes2019}. Furthermore, WHs have been extensively explored in various extended theories of gravity, such as teleparallel gravity in \cite{sahoo2018}. In addition, intriguing research on the wormhole solution for three distinct scenarios, Various fluids, including isotropic, anisotropic, and barotropic fluids, in $f(R, T)$ gravity, has been conducted in \cite{capozziello2012}. Furthermore, wormhole configurations in $ \kappa(R,T) $ gravity with linear and non-linear functions were investigated in \cite{ksh23}. The first solutions modeling compact stars in $\kappa(\mathcal{R},\mathcal{T})$-gravity for isotropic coordinates were published by Teruel et al. \cite{grp22}. Additionally, a new topic of Gravaster solution in $\kappa(\mathcal{R},\mathcal{T})$-gravity was introduced by Teruel et al. in \cite{grp24}.

The rationale for extending $f(Q)$ gravity is rooted in the history of cosmic expansion. Researchers have shown the efficacy of modified gravity models of $f(Q)$ gravity in elucidating the dynamics of the late-time universe. These frameworks are operated on the premise that non-metricity serves as the foundational geometric parameter \cite{an23}. Dynamical analysis and cosmography in $f(Q)$ gravity have been conducted in previous studies \cite{lu19}. Noether symmetry, requirements of energy, and cosmological perturbations are explored for $f(Q)$ models in \cite{li21}. $f(Q)$ theory has been used for the formulation of bouncing universe models and a range of alternative gravity theories \cite{xu20, harko18}.

Calculating the gravitational, anisotropic, hydrostatic, and modified gravitational forces is essential for evaluating the stability of models under equilibrium conditions \cite{sh20}. A traversable wormhole can only be constructed when exotic matter is present, potentially exhibiting negative energy and violating conventional energy conditions \cite{mu23}. One of the most widely used methods to quantify the total amount of energy-condition-violating matter is the "volume integral quantifier" \cite{A22}. Nandi et al. \cite{k04} further enhanced this method to precisely determine the amount of exotic matter within a specified space-time.

Wormholes have the potential to connect very huge distances—spanning billions of light-years, short distances such as meters, various points in time, or even separate universes \cite{m21}. Bueno et al. \cite{bu18} measured gravitational waves and searched for echoes of the signals of gravitational waves at black hole size \cite{mu23a,av22} presenting a possible observational technique lately. These gravitational waves are expected to correlate with the ringdown phase that appears when binary coalescences unite. Another interesting idea for observation is the proposal made by Paul et al. \cite{pa20}  involving identifying unique patterns in accretion disc \cite{ha22} to discriminate between a black hole and a wormhole geometry. Traversable wormholes are proposed as a way to access pathways to alternate or parallel universes, and as a substitute for space travel in others \cite{ta23}. From this vantage point, numerous approaches have been used to address the problem. Several writers have detected wormhole geometries in the literature \cite{de21,ca18,av22a} by investigating different types of exotic matter in the context of modified gravity theories.

The manuscript is organized as follows: Section II introduces the fundamental field equations within $f(Q)$ gravity. Section III addresses the essential conditions for wormhole construction and energy constraints. In Section IV, we delve into the traversable wormhole framework in $f(Q)$ gravity, formulating the equations of motion for a linear form of  $f(Q)$ gravity. Sections V and VI explore three wormhole solutions that comply with the Chaplygin Gas model of $f(Q)$ gravity, along with one solution aligned with the varying barotropic Fluid model. In section VII, we generate some embedding diagrams for the constructed wormhole models. In section VIII, we present the stability analysis using TOV equation for our model. Lastly, we present our conclusions in Section IX.

\section{Basic field equations in $f(Q)$ gravity}
The action of  $f(Q)$ gravity \cite{jimenez2018} is expressed as
\begin{equation}
S =\int \left(\frac{f(Q)}{16\pi} +L_m \right) \sqrt{-g} d^4x,\label{e1}
\end{equation}
where $f(Q)$ is an arbitrary function of the non-metricity $Q$, $g$
is the determinant of the metric $g_{\mu\nu}$ and $L_m$ stands for the matter Lagrangian density. The non-metricity tensor is defined as
\begin{equation}
Q_{\alpha\mu\nu} = \nabla_\alpha g_{\mu\nu} = -L^\rho_{\alpha\mu}g_{\rho\nu} - L^\rho_{\alpha\nu}g_{\rho\mu}, \label{e2}
\end{equation}
where the disformation term is written as,
\begin{equation}
L^\alpha_{\mu\nu} = \frac{1}{2}Q^\alpha_{\mu\nu} - Q^\alpha_{(\mu\nu)},\label{e3}
\end{equation}
and the traces of two independent non-metricity tensor
are written as
\begin{equation}
Q_\alpha = Q^\alpha_{\mu\mu} , \tilde{Q}_\alpha = Q^\mu_{\alpha\mu} . \label{e4}
\end{equation}
In this framework, the non-metricity scalar is defined through the contraction of $Q_{\alpha\beta\gamma}$ as
\begin{equation}
Q = -g^{\mu\nu}\left(L^\alpha_{\beta\nu}L^\beta_{\mu\alpha} - L^\beta_{\alpha\beta}L^\alpha_{\mu\nu}\right)
= -P^{\alpha\mu\nu}Q_{\alpha\mu\nu}. \label{e5}
\end{equation}
where $P_{\alpha\beta\gamma}$  is the non-metricity conjugate, and the associated tensor is expressed as
\begin{equation}
4P^\alpha_{\mu\nu} = -Q^\alpha_{\mu\nu} + 2Q^\alpha_{(\mu\nu)} - Q^\alpha g_{\mu\nu}
-\tilde{Q}^\alpha g_{\mu\nu} - \delta^\alpha_{(\mu}Q_{\nu)}. \label{e6}
\end{equation}
Variation of equation \eqref{e1} with respect to $g_{\mu\nu}$ yields the field equations 
\begin{equation}
2\sqrt{-g}\nabla_\alpha\left(\sqrt{-g}f_Q P^\alpha_{\mu\nu}\right)
+\frac{1}{2}g_{\mu\nu}f + f_Q\left(P_{\mu\alpha\beta}Q^\alpha_{\beta\nu} - 2Q_{\alpha\beta\mu}P^\alpha_{\beta\nu}\right)
= -8\pi T_{\mu\nu}, \label{e7}
\end{equation}
where for notational simplicity, we write $f_Q = f'(Q)$ and
the energy-momentum tensor $T_{\mu\nu}$ is given by
\begin{equation}
T_{\mu\nu} = -\frac{2}{\sqrt{-g}}\frac{\delta\sqrt{-g}\mathcal{L}_m}{\delta g_{\mu\nu}}. \label{e8}
\end{equation}
Varying \eqref{e1} with respect to the connection, one obtains
\begin{equation}
\nabla_\mu\nabla_\nu\left(\sqrt{-g}f_Q P^{\mu\nu}_{~~~\alpha}\right)
= 0. \label{e9}
\end{equation}

The field equations thus formulated ensure energy-momentum tensor consistency within the $f(Q)$ gravity framework. This discussion primarily aims to derive the gravitational field equations describing static and spherically symmetric spacetimes, specifically targeting wormhole geometries, as defined in equation \eqref{e7}.

\section{Basic conditions for wormholes and energy conditions}
We now examine a general spherically symmetric spacetime of a static wormhole from the Morris-Thorne class \cite{m88}. This spacetime metric can be expressed as
\begin{eqnarray}
ds^2&=&-e^{2\Phi(r)} dt^2 + \left(1 - \frac{b(r)}{r}\right)^{-1} dr^2 + r^2 d\theta^2 + r^2 \sin^2\theta d\phi^2 ,\label{e10}
\end{eqnarray}
where $\Phi(r)$ is the redshift function depending on the radial coordinate $r$ $(0 < r_0 \leq r \leq \infty)$ which remains finite everywhere to avoid any event horizon. The function 
$b(r)$, called the shape function, defines the geometry of the wormhole. For a viable wormhole structure, $b(r)$ must fulfill the following conditions:

\begin{itemize}
    \item Throat condition: $b(r_0) = r_0$ and for $r > r_0$ , the function $b(r) < r$ .
    \item Flaring out condition: $b'(r_0) < 1$, i.e., $\frac{b(r) - rb'(r)}{b^2(r)} > 0$, where $'$ denotes the derivative with respect to $r$.
    \item Asymptotically flatness condition: $\frac{b(r)}{r} \to 0$ as $r \to \infty$.
\end{itemize}

In this setup, the matter content within the wormhole is represented by an anisotropic stress-energy tensor as follows
\begin{equation}
T_\mu^\nu = (\rho+p_t) u_\mu u^\nu - P_t\, \delta_\mu^\nu + (P_r - P_t) v_\mu v^\nu.~\label{e11}
\end{equation}
where $u_\mu$ is the four-velocity, $v_\mu$ is a unit, space-like vector in the radial direction, $\rho$ is the energy density, $P_r$ and $P_t$ are the radial and tangential pressures respectively, which are functions of the radial coordinate $r$.

\subsection{Energy Conditions}
The energy conditions aim to provide a physically real matter outline, as derived from the Raychaudhuri equations. These equations describe the time evolution of the expansion scalar ($\theta$) for the congruences of timelike ($u^\mu$) and null ($\eta^\mu$) geodesics given as \cite{ar55}

\begin{equation}
\frac{d\theta}{d\tau} - \omega_{\mu\nu}\omega^{\mu\nu} + \sigma_{\mu\nu}\sigma^{\mu\nu} + \frac{1}{3}\theta^2 + R_{\mu\nu}u^\mu u^\nu = 0 ~\label{e12}
\end{equation}
\begin{equation}
\frac{d\theta}{d\tau} - \omega_{\mu\nu}\omega^{\mu\nu} + \sigma_{\mu\nu}\sigma^{\mu\nu} + \frac{1}{2}\theta^2 + R_{\mu\nu}\eta^\mu\eta^\nu = 0 \label{e13}
\end{equation}

where $\sigma^{\mu\nu}$ and $\omega_{\mu\nu}$ are respectively the shear and the rotation associated with the vector field $u^\mu$. For attractive gravity, we have $\theta < 0$. Ignoring quadratic terms, the Raychaudhuri equations give 
\begin{equation}
R_{\mu\nu}u^\mu u^\nu \geq 0~\label{e14}
\end{equation}

\begin{equation}
R_{\mu\nu}\eta^\mu\eta^\nu \geq 0 \label{e15}
\end{equation}
One of the key features of wormhole solutions in General Relativity (GR) is the violation of energy conditions (ECs). These conditions serve as criteria for ensuring the positivity of the stress-energy tensor. The four widely recognised energy conditions are the null energy condition (NEC), dominant energy condition (DEC), weak energy condition (WEC), and strong energy condition (SEC).

\begin{itemize}
    \item \textbf{N.E.C}: \(\rho + p \geq 0\), \(\rho + p_t \geq 0\)
    \item \textbf{W.E.C}: \(\rho \geq 0\), \(\rho + p \geq 0\), \(\rho + p_t \geq 0\)
    \item \textbf{D.E.C}: \(\rho \geq 0\), \(\rho - |p| \geq 0\), \(\rho - |p_t| \geq 0\)
    \item \textbf{S.E.C}: \(\rho + p \geq 0\), \(\rho + p_t \geq 0\), \(\rho + p + 2p_t \geq 0\)
\end{itemize}

\section{Wormhole geometries in $f(Q)$ gravity}
In this part, we talk about the different types of wormhole solutions with self-stability. The form of the trace of the non-metricity tensor $Q$ for the wormhole metric in \eqref{e10} is given by
\begin{equation}
Q = -\frac{2}{r} \left(1 - \frac{b(r)}{r}\right) \left(2\phi' + \frac{1}{r}\right).~\label{e16}
\end{equation}

By putting equations \eqref{e10} and \eqref{e11} into equation \eqref{e7}, we obtain the following field equations

\begin{equation}
\rho =\left[ \frac{1}{r} \left(\frac{1}{r} - \frac{rb'(r) + b(r)}{r^2} + 2\phi'(r) \left(1 - \frac{b(r)}{r}\right)\right)\right] f_Q + \frac{2}{r} \left(1 - \frac{b(r)}{r}\right) f'_Q + \frac{f}{2}.~\label{e17}
\end{equation}

\begin{equation}
P_r = -\left[\frac{2}{r} \left(1 - \frac{b(r)}{r}\right) \left(2\phi'(r) + \frac{1}{r}\right) - \frac{1}{r^2}\right] f_Q - \frac{f}{2}.~\label{e18}
\end{equation}

\begin{eqnarray}
P_t&=&-\left[\left(1 - \frac{b(r)}{r}\right) \left(\frac{1}{r} + \phi'(r) (3 + r\phi'(r)) + r\phi''(r)\right) - \frac{rb'(r) + b(r)}{2r^2} \left(1 + r\phi'(r)\right)\right] \frac{f_Q}{r}\nonumber \\
    &-& \frac{1}{r}\left(1 - \frac{b(r)}{r}\right) (1 + r\phi'(r)) \dot{f_Q}- \frac{f}{2}.~\label{e19}
\end{eqnarray}
where $f \equiv f(Q)$, $f_Q = \frac{df(Q)}{dQ}$ and $f_{QQ} = \frac{d^2 f(Q)}{dQ^2}$. Finally, we have three independent equations \eqref{e17}–\eqref{e19} for our six unknown quantities, i.e., $\rho(r)$, $P(r)$, $P_t(r)$, $\Phi(r)$, $b(r)$ and $f(Q)$. Thus, the above system of equations is under-determined, and it is possible to adopt different strategies to construct wormhole solutions. Here, we will focus on a particularly interesting case that follows a constant redshift function, $\Phi=Constant$. With this assumption, one can simplify the calculations considerably and provide interesting exact wormhole solutions.

\subsection{Wormhole solutions in linear $f(Q)$ model}
Here, we consider a linear form of the $f(Q)$ gravity model and explore wormhole solutions. Field equations will be obtained by using this form of $f(Q)$, and subsequently, some particular dark energy models will be discussed. We consider the linear functional form of $Q$ as \cite{ZH21},
\begin{equation}
f(Q) = \alpha\, Q ~\label{e20}
\end{equation}
where $\alpha$ is a constant.A linear form of the $f(Q)$ model reproduces the equations of GR without curvature or torsion, but through non-metricity. This is the symmetric teleparallel equivalent of GR. This gives a geometrically different but comparable formulation of gravity. Simpler second-order field equations result from a linear $f(Q)$ model.  They are therefore easier to handle for analytical and numerical research, especially when we are dealing with a complicated wormhole configuration. Under the right conditions, such models can describe the late cosmic acceleration and can mimic dark energy. A linear $f(Q)$ theory avoids ghost degrees of freedom and preserves some desired characteristics, such as gauge invariance. Since our study involves wormhole configurations in the background of a modified gravity and matter in the form of dark energy, the project is complex at the analytical level. So it is better that we keep it simple and consider a linear form of the modified gravity. However, a non-linear form can also be considered without changing the basic physical implication. The redshift function $\Phi(r)$ in Eq.~\eqref{e10} must be finite and non-vanishing at the throat $r_0$. Thus, $\Phi(r) = \text{constant}$ can be considered to achieve de Sitter and anti-de Sitter asymptotic behaviour. Therefore, the field equations in Eqs.~\eqref{e17} to \eqref{e19} are transformed into,

\begin{equation}
-\frac{b'(r)}{r^2}\alpha = \rho~\label{e21}
\end{equation}

\begin{equation}
\frac{b(r)}{r^3}\alpha = P_r~\label{e22}
\end{equation}

\begin{equation}
\left(\frac{b'(r)}{2r^2} - \frac{b(r)}{2r^3}\right)\alpha = P_t~\label{e23}
\end{equation}
Now, in the next step, we will discuss three different cases of WH solutions for some dark energy equation of states.

\section{Wormhole Models with Chaplygin gas in linear $f(Q)$ model}
In this part, we present wormhole configurations using two Chaplygin gas models. We explore wormhole geometries in Varying Chaplygin gas \cite{Elizalde18} and Generalized Chaplygin gas \cite{BENTO02} models.

\subsection{\textbf{Model with Varying Chaplygin gas}}
Here we consider the varying Chaplygin gas (VCG) model \cite{Elizalde18}. to explore wormhole geometries. The radial pressure \eqref{e22} takes the following form \cite{Elizalde18}
\begin{equation}
p_{r}=-\frac{Bb(r)^{u}}{\rho^{a}}~\label{e24}
\end{equation}
where we preserve the $b(r)$ notation to denote that parametrized shape function and $B$, $u$, and $a$ are constants.An intriguing addition to the original Chaplygin gas framework, the Varying Chaplygin Gas model provides a more adaptable and dynamically rich method of simulating dark energy. Although dark matter and dark energy are united by a single equation of state in the generalized Chaplygin gas, it may not be able to adequately account for the changing character of cosmic acceleration over time. To better describe the universe's shift from matter domination to late-time acceleration, the VCG model adds a dynamical degree of freedom by permitting the Chaplygin gas parameters, usually the equation of state parameter $B$, to change with time or scale factor. Improved compatibility with observational data, including cosmic microwave background anisotropies, supernovae luminosity distances, and large-scale structure development, is made possible by this modification. Additionally, the VCG can provide information about how dark sectors interact, enabling richer phenomenology and possible connections to higher-dimensional physics or scalar field theories. Consequently, the VCG is a viable option for providing a more accurate and consistent description of dark energy based on observations.

\begin{figure*}
\centering
 \includegraphics[width=6.5cm,height=5cm]{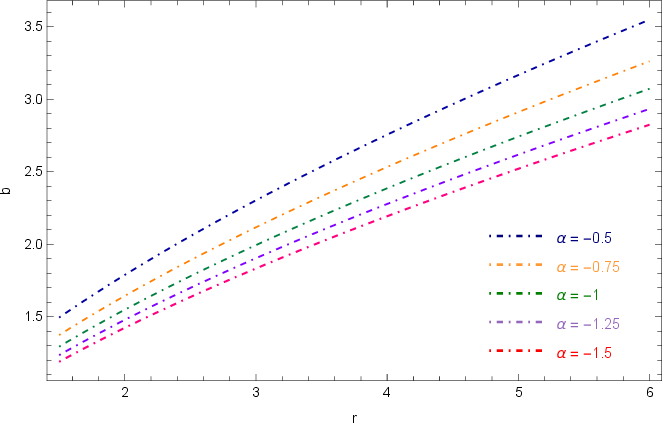} \hspace{1em}\includegraphics[width=6.5cm,height=5cm]{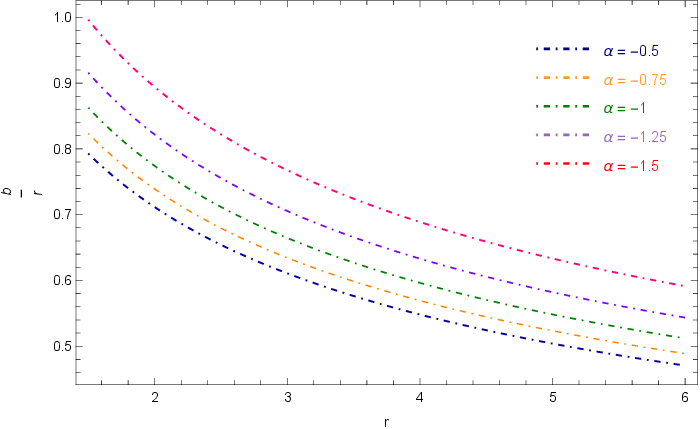}\\
    \includegraphics[width=6.5cm,height=5cm]{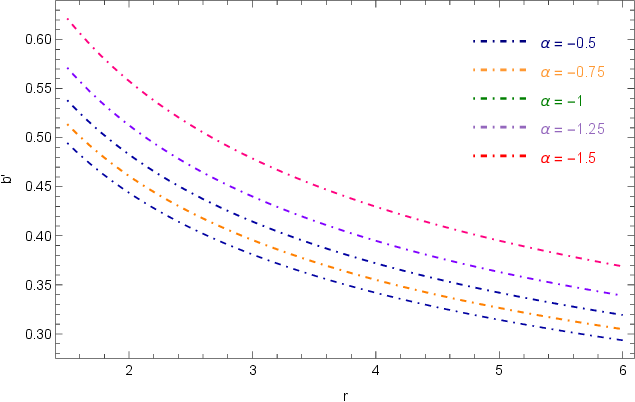} \hspace{1em}
    \includegraphics[width=6.5cm,height=5cm]{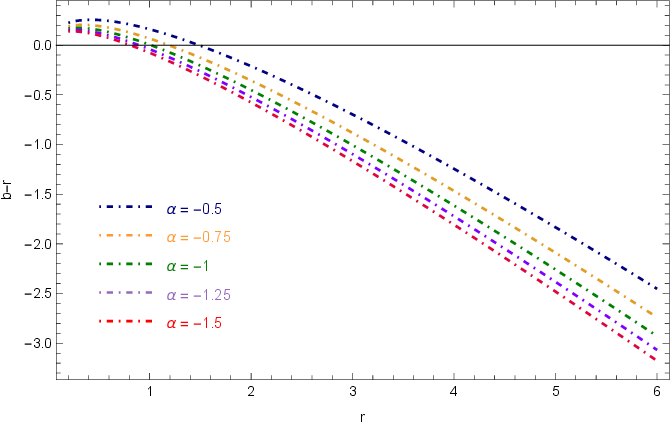}
    \caption{Characteristic of the shape functions for a wormhole with the following conditions: Flaring out condition: $ b'(r) < 1$; Throat condition: $b(r) - r < 0$; Asymptotically flatness condition: $\frac{b(r)}{r} \rightarrow 0$ as $ r \rightarrow \infty$ with varying \textbf{r}, for a = 0.05; u = -4; B = 1; c = 0.09  (for WH1).} \label{fig1}   
\end{figure*}

Since the matter in a wormhole has a radial pressure calculated via Eq.\eqref{e24}, the wormhole's shape function will look like
\begin{equation}
b(r)=\left(-\frac{(a-u+1) \left(\frac{(-1)^{1/a} \alpha ^{-\frac{1}{a}-1} B^{1/a} r^{\frac{3}{a}+3}}{\frac{3}{a}+3}-c\right)}{a}\right)^{\frac{a}{a-u+1}}~\label{e25}
\end{equation}
where $c$ is a constant. This particular solution has been obtained for $a = 0.05 ; u = -4 ; B = 1 ; c = 0.09$. The graphical behaviour of the shape function in Fig.(\ref{fig1}) indicates that a solution describing a wormhole satisfying all conditions required for traversal exists. The asymptotic flatness condition, i.e. $b(r)/r \to 0$ for $r \to \infty$, is retained in form \eqref{e25}. In Fig.(\ref{fig1}), the quantities $b(r)$, $b(r)/r$, $b'(r)$, and $b(r)-r$ are depicted by varying the radial component $r$, for $u=-4$. It is observed that $b(r) - r$ intersects the $r$-axis in Fig.~\ref{fig1}. It is noted that for a stable wormhole, the shape function $b(r)$ needs to comply with flaring, throat, and asymptotic conditions. Profiles in Fig.(\ref{fig1}) show that the shape function satisfies all the conditions required for a stable wormhole.

\begin{figure*}
    \centering
    \includegraphics[width=6.5cm,height=5cm]{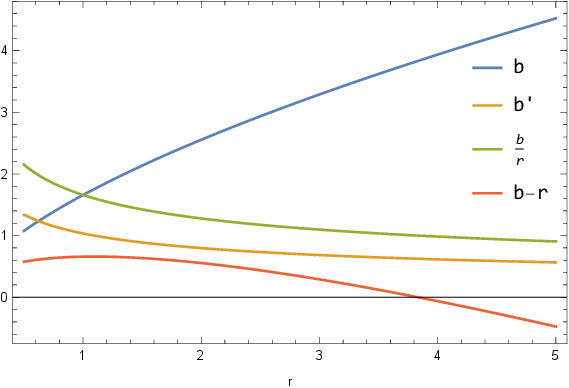} \hspace{1em}
    \includegraphics[width=6.5cm,height=5cm]{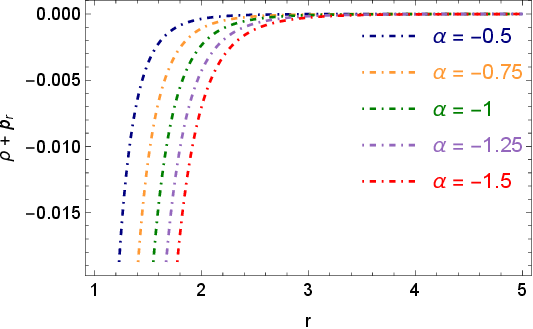}
    \caption{Profile of the shape function $b(r)$ and Null Energy Condition (NEC) with varying \textbf{r}, for a = 0.05 ; u = -4 ; B = 1 ; c = 0.09; $c_1=0.09$ (for WH1).}
    \label{fig2}
\end{figure*}

Energy conditions, as far as we know, are the greatest geometrical instrument to assess the self-stability of cosmological models. We used this method to test our models. Additionally, We have assumed that $f(Q)$ has a linear functional form. Thus, our models will continue to use the standard energy conversion conditions.\\
Using \eqref{e25}, we can now rewrite the radial pressure $p_{r}$ from \eqref{e22}, the lateral pressure $P_{t}$ from \eqref{e23}, and the energy density from equations \eqref{e21} to \eqref{e24} as follows.

\begin{eqnarray}
\rho=(-1)^{1/a} \alpha ^{-1/a} B^{1/a} r^{3/a} \left(-\frac{(a-u+1) \left(\frac{(-1)^{1/a} \alpha ^{-\frac{1}{a}-1} B^{1/a} r^{\frac{3}{a}+3}}{\frac{3}{a}+3}-c\right)}{a}\right)^{\frac{a}{a-u+1}-1}~\label{e26}
\end{eqnarray}
\begin{eqnarray}
p_r &=&\frac{\alpha  \left(-\frac{(a-u+1) \left(\frac{(-1)^{1/a} \alpha ^{-\frac{1}{a}-1} B^{1/a} r^{\frac{3}{a}+3}}{\frac{3}{a}+3}-c_1\right)}{a}\right)^{\frac{a}{a-u+1}}}{r^3},~\label{e27}
\end{eqnarray}
\begin{eqnarray}
p_t &=& \frac{\alpha}{2}  \left(-\frac{(a-u+1)\left(\frac{(-1)^{1/a} a \alpha ^{-\frac{a+1}{a}} B^{1/a} r^{\frac{3}{a}+3}}{3 (a+1)}-c_1\right)}{a}\right)^{\frac{a}{a-u+1}} \nonumber\\
    &\times& \left(-\frac{1}{r^3}+\frac{3}{(a-u+1) \left(\frac{r^3}{a+1}-\frac{3 (-1)^{-1/a} c_1 \alpha ^{\frac{1}{a}+1} B^{-1/a} r^{-3/a}}{a}\right)}\right)
.~\label{e28}
\end{eqnarray}

\begin{figure*}
\centering
\includegraphics[width=7.7cm,height=6.5cm]{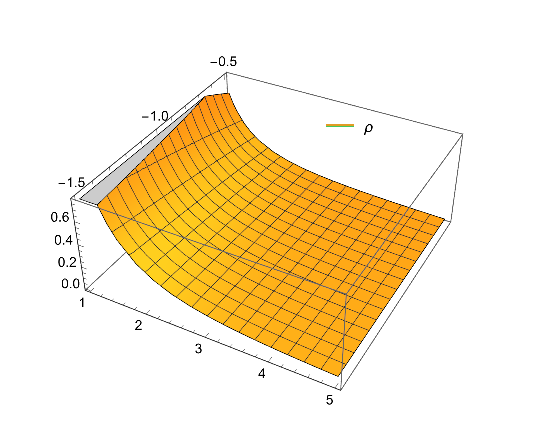}~~\includegraphics[width=7.7cm,height=6.5cm]{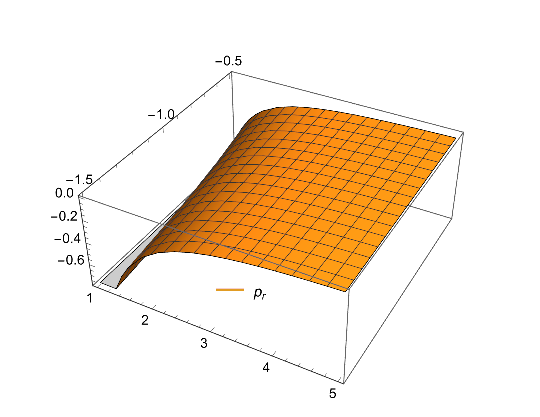}
\caption{Profile of energy density, surface pressure with respect to $r$ and $\alpha$ with $a = 0.05; u = -4; B = 1;c,c_1=0.09$ for WH1.} \label{fig3}
\end{figure*}

The behavior of the energy conditions is visible in Fig~\ref{fig2} and Fig~\ref{fig3}. Fig~\ref{fig3}  indicates the energy density $\rho$ is positive across the spacetime. The radial pressure in Fig.\ref{fig2} and Fig.\ref{fig3} shows that the NEC (Null Energy Condition) for the radial pressure is violated, meaning $\rho + p_r \le 0$ and $p_r \le 0$. These energy conditions are plotted in Fig~\ref{fig2} and Fig~\ref{fig3}, indicating that the solution satisfies all energy conditions. The violation of NEC may confirm the presence of exotic matter at the throat of the wormhole.


\subsection{\textbf{Model with Generalized Chaplygin gas}}
In this section, we will look into the generalized Chaplygin gas (GCG) model \cite{BENTO02} whose equation of state is given by
\begin{eqnarray}
p_r=-\frac{B}{\rho ^a}~\label{e29}
\end{eqnarray}
where $B$ and $a$ are constants.It exhibits the characteristics of cold dark matter in the early stages and dark energy in the later stages. Thus, dark energy and dark matter can be successfully interpolated by GCG. One may elegantly combine both elements with a single fluid. Cosmic acceleration is naturally driven by the GCG's negative pressure, as shown in CMB and supernova data. GCG models can fit a variety of data like Baryon Acoustic Oscillations (BAO), Cosmic Microwave Background (CMB), and Type Ia Supernovae, particularly when $\alpha<1$.  According to certain research, under specific priors, it matches the $\Lambda$CDM model rather well. In higher-dimensional theories, Nambu-Goto branes give rise to the original Chaplygin gas. Because of this, it has a geometric genesis, which is desirable when trying to relate high-energy physics to cosmology. In contrast to certain scalar field dark energy models, GCG does not require crossing the phantom divide (where equation of state parameter $w<-1$). It is a stable model with appropriate physical features that does not have ghosts.

\begin{figure*}
\centering
\includegraphics[width=7.3cm,height=5cm]{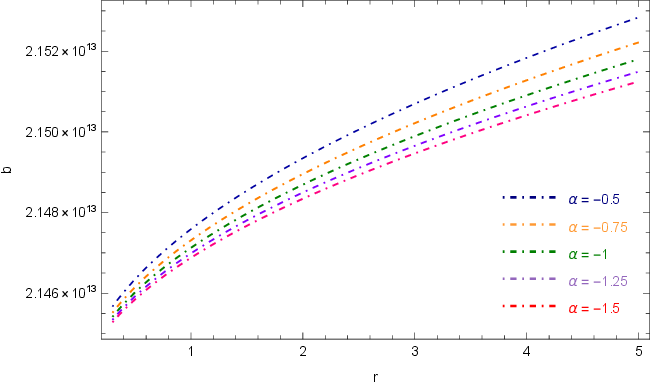}~~\includegraphics[width=7.3cm,height=5cm]{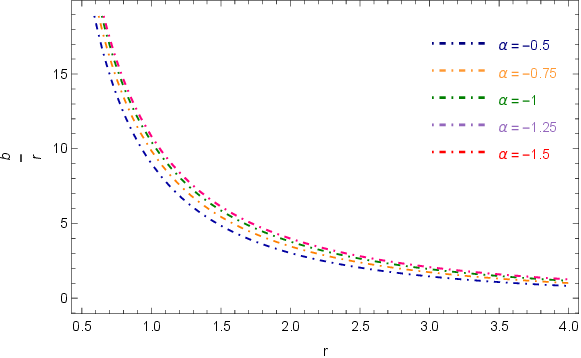}
\caption{\textbf{Profile of the shape function $b(r)$ and Asymptotically flatness condition:} \( \frac{b(r)}{r} \rightarrow 0 \) as \( r \rightarrow \infty \) with varying \textbf{r}, for $a = -1.2 ; B = 1; c_1=10$ for WH2.}\label{fig4}
\end{figure*}

\begin{figure*}
\centering
\includegraphics[width=7.3cm,height=5cm]{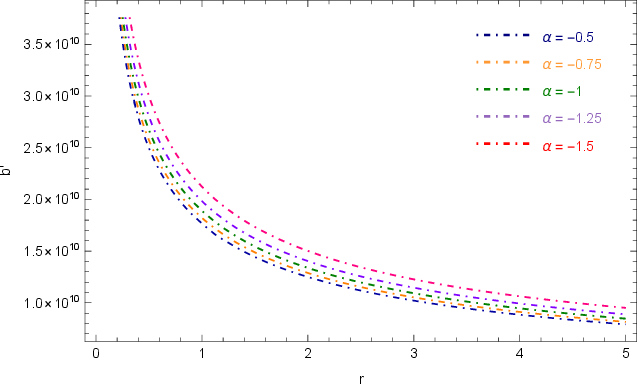}~~\includegraphics[width=7.3cm,height=5cm]{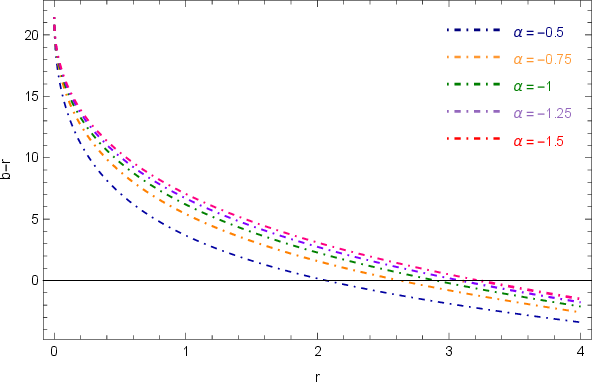}
\caption{\textbf{Profile of the shape function $b(r)$ for a wormhole with the following conditions:}
\textbf{Flaring out condition:} \( b'(r) < 1 \)\,;\,\textbf{Throat condition:} \( b(r) - r < 0 \)\,
with varying \textbf{r}, for ( a = -1.2 ; B = 1; $c_1= 10$ ) (for WH2).}\label{fig5}
\end{figure*}

For this radial pressure, the wormhole's shape function becomes,
\begin{eqnarray}
 b(r)=\left[\left(-\frac{1}{a}-1\right) \left(-\frac{(-1)^{1/a} \alpha ^{-\frac{1}{a}-1} B^{1/a} r^{\frac{3}{a}+3}}{\frac{3}{a}+3}-c_1\right)\right]^{\frac{a}{a+1}}~\label{e30}
\end{eqnarray}
where $c_1$ is a constant. We have wormhole solutions with $\rho$, $Pr$, and $P_t$. This solution has been obtained for $a = -1.2; B = 1; c_1 = 10$. From the graphical behavior of the shape function in Fig.(\ref{fig4}), we can see a solution describing a wormhole satisfying all conditions required. From \eqref{e30} we retained the asymptotic flatness condition, i.e. $b(r)/r \to 0$ for $r \to \infty$. In Fig. \eqref{fig4} and \eqref{fig5}, we obtain the quantities $b(r)$, $b(r)/r$, $b'(r)$ and $b(r)-r$ by varying the radial component $r$, for $a =-1.2$. We can see that $b(r) - r$ cuts the $r = 2$ -axis in Fig.~\ref{fig5}. Note that, for a stable wormhole, the shape function $b(r)$ needs to obey flaring, throat, and asymptotic conditions. Profiles in Fig.\ref{fig4} and \ref{fig5} show that the shape function satisfied all the conditions required for a stable wormhole.

\begin{figure*}
\centering
\includegraphics[width=6.5cm,height=6cm]{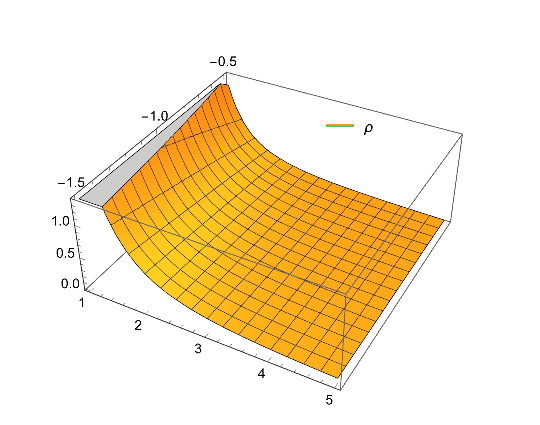}~~\includegraphics[width=6cm,height=5cm]{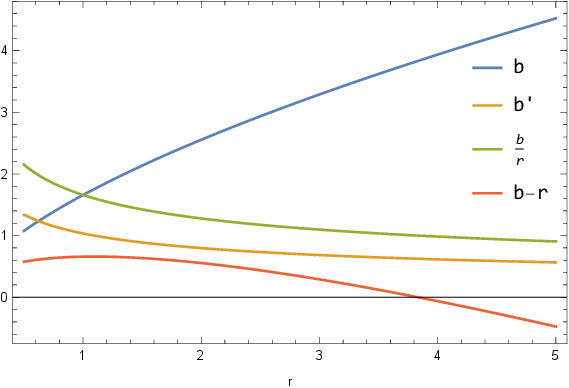}
\caption{Profile of the shape function $b(r)$  and Null Energy Condition (NEC) ~
with varying \textbf{r}, for $a = 0.05 ; u = -4 ; B = 1 ; c_1 = 0.09$ (for WH2).} \label{fig6}
\end{figure*}

\begin{figure*}
\centering
~~\includegraphics[width=6.5cm,height=6cm]{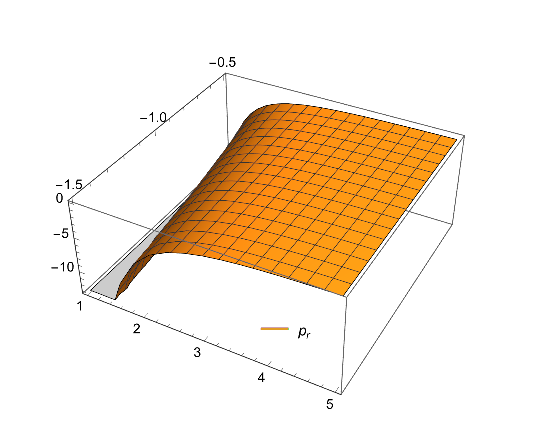}~~\includegraphics[width=6.5cm,height=6cm]{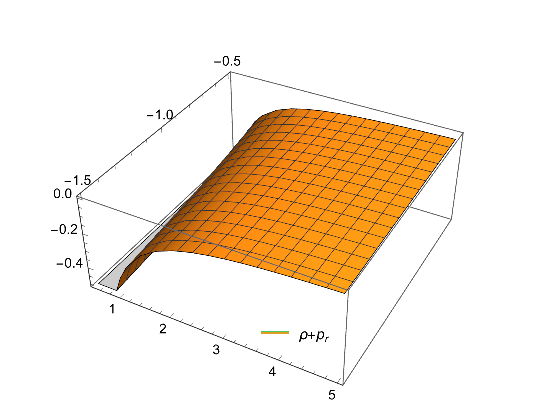}
\caption{Variations of surface pressure and Null Energy Condition (NEC)} \label{fig7}
\end{figure*}

Using \eqref{e30}, we can now rewrite the radial pressure $P_{r}$ from \eqref{e22}, lateral pressure $P_{t}$ from \eqref{e23}, and energy density from equations from \eqref{e21} through \eqref{e29} as follows.

\begin{eqnarray}
\rho=-\frac{(-1)^{\frac{1}{a}} \left(-\frac{1}{a}-1\right) a \alpha ^{\frac{-1}{a}} B^{\frac{1}{a}} r^{\frac{1}{a}} \left(\left(-\frac{1}{a}-1\right) \left(-\frac{(-1)^{\frac{1}{a}} \alpha ^{-\frac{1}{a}-1} B^{\frac{1}{a}} r^{\frac{3}{a}+3}}{\frac{3}{a}+3}-c_1\right)\right){}^{\frac{a}{a+1}-1}}{a+1}~\label{e31}
\end{eqnarray}
\begin{eqnarray}
P_r=\frac{\alpha  \left(\left(-\frac{1}{a}-1\right) \left(-\frac{(-1)^{\frac{1}{a}} \alpha ^{-\frac{1}{a}-1} B^{\frac{1}{a}} r^{\frac{3}{a}+3}}{\frac{3}{a}+3}-c_1\right)\right)^{\frac{a}{a+1}}}{r^3}~\label{e32}
\end{eqnarray}
\begin{align}
P_t=\frac{
\alpha \left(2 (-1)^{1/a} a B^{1/a} r^{\frac{3}{a}+3} 
    + 3 (a+1) c \alpha^{\frac{1}{a}+1}\right)
    \left(-\frac{1}{3} e^{\frac{i \pi}{a}} \alpha^{-\frac{a+1}{a}} 
    B^{1/a} r^{\frac{3}{a}+3} + \frac{c}{a} + c\right)^{\frac{a}{a+1}}
}{
    2 (-1)^{1/a} a B^{1/a} r^{\frac{3}{a}+6} 
    - 6 (a+1) c r^3 \alpha^{\frac{1}{a}+1}
}~\label{e33}
\end{align}
Figure \ref{fig6} shows the behaviour of energy conditions (ECs). We have set the settings to a = -1.2; B = 1; c = 10 to validate the EC profiles. Energy density, as is generally known, has to be positive everywhere, however, there may be exceptions. Moreover, $P_r < 0 $ and by this explanation we can say that $P_t > 0$ was observed in Fig.(\ref{fig6}). Furthermore, we've shown the NEC graph in Figure \ref{fig6}. The peculiar material at the wormhole throat, which may be required for the geometry of wormholes, is demonstrated by the NEC violation for different parameter values. We found that the NEC in terms of both pressures remains true even in the locations where the NEC in terms of $P_r$ is broken. Furthermore, we find that $\rho>0$ is fulfilled for the areas and range of the parameters investigated (see fig.(\ref{fig7})).


\section{Wormhole Models with Varying Barotropic fluid}
Exact wormhole solutions from the assumption that a Varying barotropic fluid (VBF) \cite{Elizalde18} can describe the radial pressure of the matter content of the wormhole will be considered in this section.
\begin{equation}
p_r = -\omega\,b(r)^u \,\rho~\label{e39}
\end{equation}
This suggests that we consider a parametrization of an evolving equation of state affected by the shape function $b(r)$.A flexible framework for investigating the dynamics of dark energy and its role in cosmic acceleration is provided by the varying barotropic fluid model.  The VBF technique permits the barotropic index $\omega$ to change with time or the scale factor, allowing the fluid to adjust its behavior over different cosmological epochs, in contrast to conventional barotropic fluids with a constant equation of state $ p=\omega\rho$.  While still being consistent with observable data, this time-dependence depicts the change from a decelerated, matter-dominated universe to the current accelerating expansion. Additionally, the VBF model can naturally handle crossing of the phantom divide without causing instabilities, and thus avoids some of the fine-tuning problems found in constant-$\omega$ models. Furthermore, by presenting rich cosmic dynamics that can be limited by supernova, CMB, and structure growth measurements, it offers a phenomenological link between quintessence-like behavior and more exotic dark energy possibilities. Because of this, the varying barotropic fluid is a strong contender to represent the intricate development of dark energy.

\begin{figure*}
\centering
\includegraphics[width=6.5cm,height=5cm]{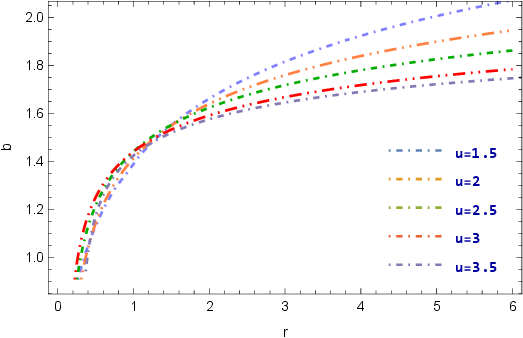}~~\includegraphics[width=6.5cm,height=5cm]{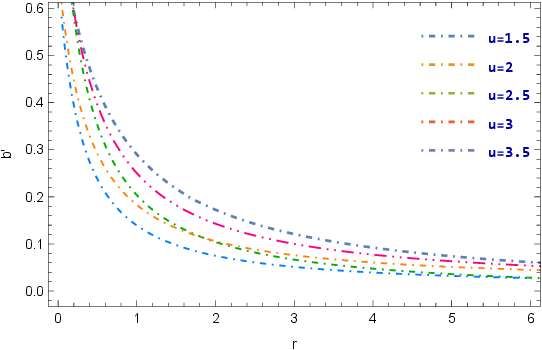}
\caption{\textbf{Profile of the shape functions and  Flaring out condition where } $ b'(r) < 1 $ for $(\omega=-2~ and~ c_1=1)$} \label{fig10}
\end{figure*}

\begin{figure*}
\centering
\includegraphics[width=6.5cm,height=5cm]{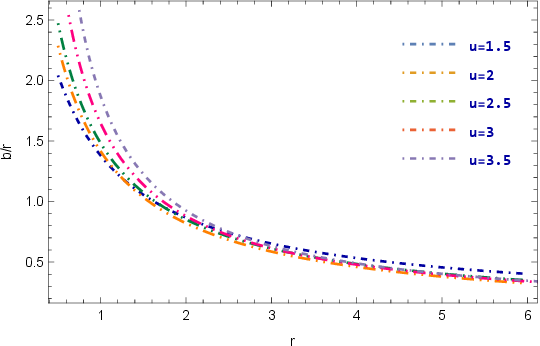}~~\includegraphics[width=6.5cm,height=5cm]{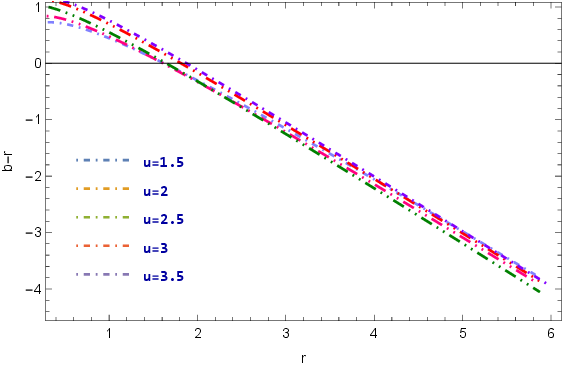}
\caption{\textbf{Characteristics of shape function $b(r)$ for a wormhole with the following conditions:}
\textbf{Throat condition:} \( b(r) - r < 0 \);\textbf{Asymptotically flatness condition:} \( \frac{b(r)}{r} \rightarrow 0 \) as \( r \rightarrow \infty \)
with varying \textbf{r}, for $\omega=-2 ; c_1 = 1$ (for WH3).} \label{fig11}
\end{figure*}

\begin{figure*}
\centering
\includegraphics[width=6cm,height=5cm]{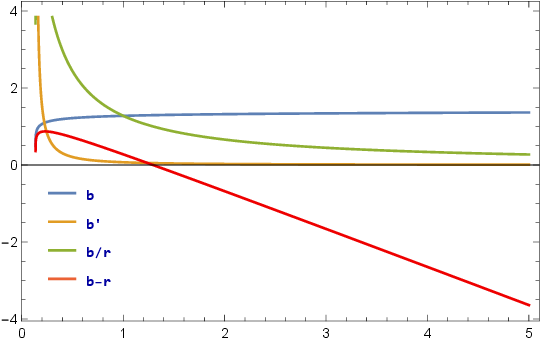}~~\includegraphics[width=6.5cm,height=6cm]{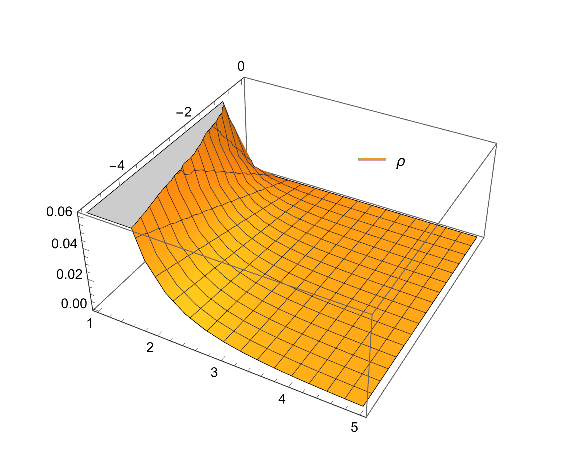}
\caption{\textbf{Profile of Shape function and energy density w.r.t. r and $u$ with $c_1=1,\omega=2,\alpha=-0.5$ for WH3.}}\label{fig12}
\end{figure*}

Nevertheless, the shape function defining a wormhole adopts the specific form as follows
\begin{equation}
b(r)=\left(-u \left(-\frac{\log (r)}{\omega }-c_1\right)\right)^{1/u}~\label{e40}
\end{equation}
where $c_1$ is the constant. Then, using the given shape function, one may derive the explicit forms of $P_r$ and $\rho$. A more generalised scenario may be numerically analyzed, as seen below for the wormhole solution with $\omega = 2; c_1 = 1$. Furthermore, the same plots show the behavior of the radial and lateral equation of state parameters, suggesting that wormhole creation is feasible as long as the radial equation of state parameters is phantom-like.
Our answer describes a wormhole that satisfies all the requirements necessary for the wormhole. The asymptotic flatness requirement, that is, $b(r)/r \to 0$ for $r \to \infty$, was maintained in eqn. \eqref{e40}. The numbers $b(r)$, $b(r)/r$, $b'(r)$, and $b(r)-r$ curve are shown in Fig. \eqref{fig10},\eqref{fig11}and \eqref{fig12} for a radial component $r$ of $u = 1$. In Fig.~\ref{fig11}, it is evident that $b(r) - r$ intersects the $r=1.2$-axis. It must be kept in mind that the shape function $b(r)$ must abide by asymptotical constraints, such as throat and flaring criteria, for a stable wormhole. The profiles presented in Fig.\ref{fig12} demonstrate that all the requirements for a stable wormhole were met by the shape function. For $r_0 = 1.2$, the neck of the wormhole is reached, while $b'(r_0)<1$. 

\begin{figure*}
\centering
\includegraphics[width=6.5cm,height=5cm]{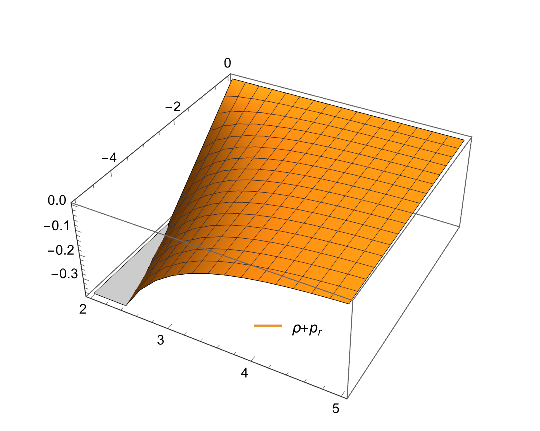}~~\includegraphics[width=6.5cm,height=5cm]{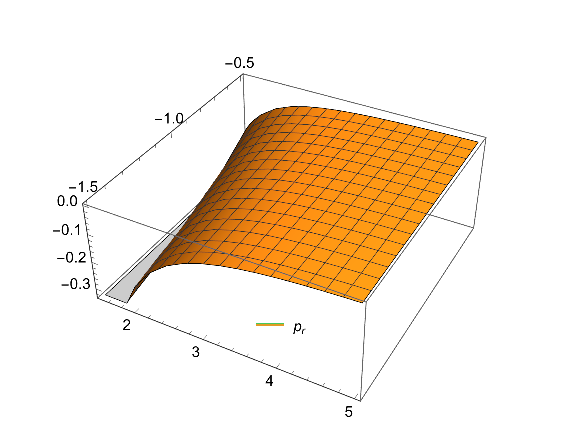}
\caption{\textbf{Characteristics of NEC and Radial pressure w.r.t. r and $u$ with $c_1=1,\omega=2,\alpha=-0.5$ for WH3.}} \label{fig13}
\end{figure*}

\begin{figure*}
\centering
~~\includegraphics[width=8cm,height=6cm]{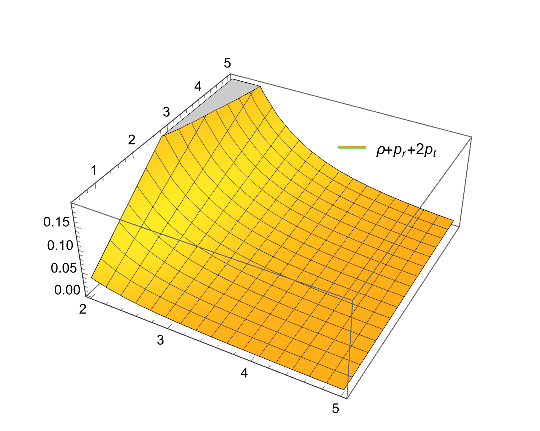}
\caption{\textbf{Characteristics of WEC and $u$ with $c_1=1,\omega=2,\alpha=-0.5$ for WH3.}} \label{fig14}
\end{figure*}

Now using \eqref{e22} and \eqref{e40} and lateral pressure $P_{t}$ from \eqref{e23}, and energy density from equations from  \eqref{e21} we have
\begin{equation}
 \rho=-\frac{\alpha  \left(-u \left(-\frac{\log (r)}{\omega }-c_1\right)\right)^{\frac{1}{u}-1}}{r^3 \omega } ~\label{e41}
\end{equation}
\begin{equation}
  p_r=\frac{\alpha  \left(-u \left(-\frac{\log (r)}{\omega }-c_1\right)\right){}^{1/u}}{r^3}~\label{e42}
\end{equation}
\begin{equation}
p_t=-\frac{\alpha  (u \log (r)+c_1 u \omega -1) \left(u \left(\frac{\log (r)}{\omega }+c_1\right)\right){}^{\frac{1}{u}-1}}{2 r^3 \omega }~\label{e43}
\end{equation}
The NEC will be violated in the neck of a wormhole of this kind, but it will remain valid and far from the throat, as can be shown from the graphical behaviour of the NEC in terms of $P_r$, as illustrated in Fig.\eqref{fig13}. From Figs. \eqref{fig13}, and \eqref{fig14},we can see the behaviour of the energy conditions. In Fig.\eqref{fig12}, we can see that the energy density is positive throughout spacetime. From Fig.\eqref{fig13} and \eqref{fig14}, we can see that the radial pressure and Energy condition $\rho +p_r +2p_t$ is valid throughout spacetime, and we can see that the NEC is violated in that region, i.e., $\rho +p_r\,<0$. From this behaviour of the energy density, radial pressure, and NEC, we can say that it satisfies all the energy conditions except NEC, which may be used for wormhole geometry.


\section{Embedding Diagram}

In this section, embedding diagrams which will prove to be useful to visualize the wormhole spacetime in \eqref{e10}, will be presented. The geometry of the Morris-Thorne wormhole \cite{MT88} at the equatorial plane ($\phi = \pi /2$) on a slice of constant time $t$ is given by
\begin{eqnarray}
ds^{2}=\left[1-\frac{b}{r}\right]^{-1}dr^{2} +r^{2}d\theta^{2} \label{e44}
\end{eqnarray} 
and embed on the 3-dimensional cylinder of the form
\begin{eqnarray}
ds^2 = dr^2+dz^2+r^2 d\theta^2. \label{e45}
\end{eqnarray}
On comparing \eqref{e44} and \eqref{e45} we get,
\begin{equation}
1+\left({dz \over dr} \right)^2={1 \over 1-b/r}  ~~~\text{or}~~~\left({dz \over dr} \right)^2={1 \over 1-b/r}-1\nonumber\\={1 \over r/b-1}.
\end{equation}
Now the embedding surface can be determined as
\begin{eqnarray}
z(r) = \pm \int_{r_{_0}}^r {dr \over \sqrt{r/b(r)-1}}.\label{e46}
\end{eqnarray}
In \eqref{e46}, we put the respective shape functions $b(r)$, and we get the redshift function solutions of the three wormhole models to construct the embedding surfaces. Fig.\ref{fig13},\ref{fig14} illustrates the corresponding embedding diagrams for the models WH1, WH2, and WH3, each associated with a different wormhole configuration.

\begin{figure*}
\centering
\includegraphics[width=6.5cm,height=5cm]{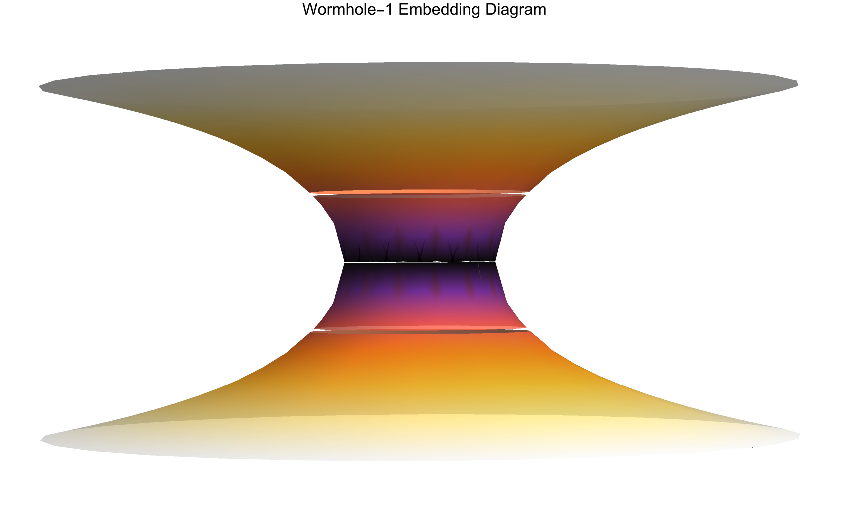}~~\includegraphics[width=6.5cm,height=5cm]{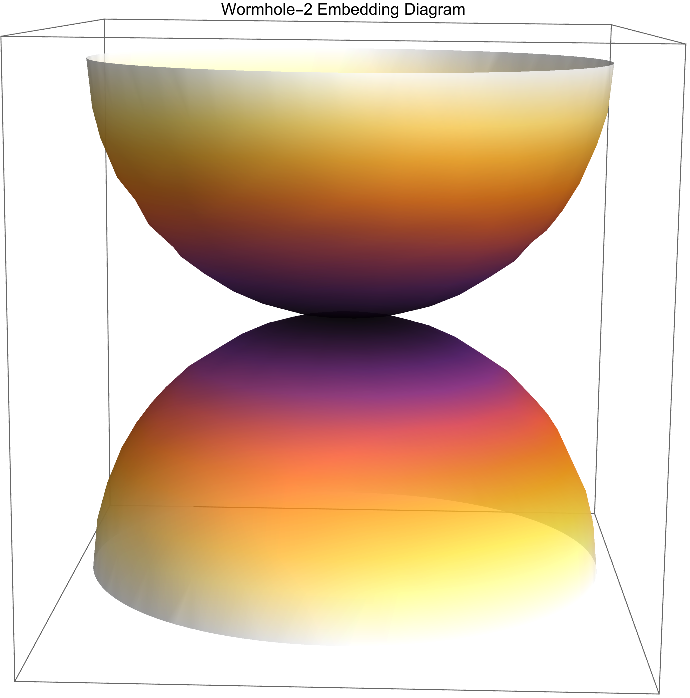}
\caption{The wormhole is embedded in a three-dimensional space-time. In the plot, $b_0$ is chosen to be $ 1M$ for WH1 and $ b_0=2M$ for WH2.} \label{fig15}
\end{figure*}

\begin{figure*}
\centering
\includegraphics[width=6.5cm,height=5cm]{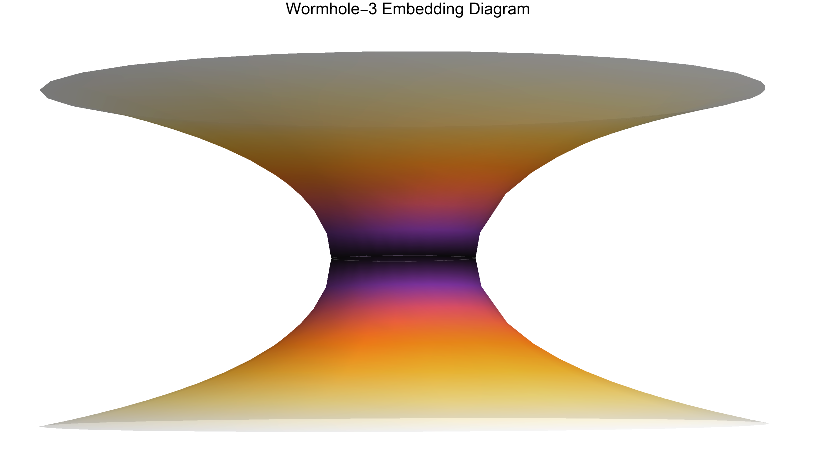}
\caption{A three-dimensional space-time embeds the wormhole, represented by the given plot with specifications: $(c=0;u=1;\omega =-1;)$. In the plot, we set $b_0 = 2M$.} \label{fig16}
\end{figure*}

The solutions for $b(r)$, evaluated for three wormhole models in Eq. \eqref{e46}, were used to draw the embedding surfaces. This is a special case of Eq. \eqref{e44}, where the coordinate $z$ is related to $r$. The "redshift function'' vanishes ($\Phi = constant$). In Fig.\ref{fig13},\ref{fig14}, the embedding surfaces for wormhole1, wormhole2, and wormhole3 are shown concerning three models. The solution is very difficult to calculate, which is why $z(r)$ values of free parameters were then calculated to discuss the energy conditions for respective models (see figs.(\ref{fig15}), (\ref{fig16})).

\section{Stability from the TOV Equation}
To study the equilibrium condition of the formation of the wormhole in $ f(Q) $ gravity \cite{jimenez2018}, we use the generalised Tolman-Oppenheimer-Volkoff (TOV) equation \cite{JR374, RC364}. Although it was initially expressed for stellar interiors, the framework can be stretched to analyze the equilibrium structure of wormhole spacetimes, mainly in modified gravity models like $f(Q)$ gravity.
Generally, the distribution of the matter is anisotropic, suggesting that the radial pressure ($p_r$) and the tangential pressure ($p_t$) are unequal. For this type of condition, the equilibrium condition is preserved by the three combined forces: hydrostatic force, gravitational force, and the force resulting from anisotropy in the pressure distribution.
More generally, the TOV equation for an anisotropic fluid can be expressed as:
\begin{equation}
\frac{dp_r}{dr} + \frac{\phi'}{2} (\rho + p_r) + \frac{2}{r} (p_r - p_t) = 0,\label{s1}
\end{equation}
where $ \phi $ is the gravitational redshift function, $\phi'$ is the derivative of the gravitational redshift function $\phi$ with respect to $r$ , $ \rho $ is the energy density, $ p_r $ is the radial pressure, and $ p_t $ is the tangential pressure.

This equation characterises the equilibrium of 3 different forces:
\begin{itemize}
    \item The hydrostatic force written as $ F_H = -\frac{dp_r}{dr} $,
    \item The gravitational force written as $ F_G = -\frac{\phi'}{2} (\rho + p_r) $,
    \item The anisotropic force written as  $F_A = \frac{2}{r} (p_t - p_r) $.
\end{itemize}

Thus, the stability equilibrium condition \eqref{s1} can be written as,
\begin{equation}
F_H + F_G + F_A = 0.
\end{equation}
For our choice of the redshift $\phi$ as a constant, we have $ F_G = -\frac{\phi'}{2} (\rho + p_r) $, which vanishes, i.e., the gravitational force does not impact the stability of our model. This simplifies our condition as, 
\begin{equation*}
    F_H + F_A = 0.
\end{equation*}

\begin{figure*}
\centering
 \includegraphics[width=6.5cm,height=5cm]{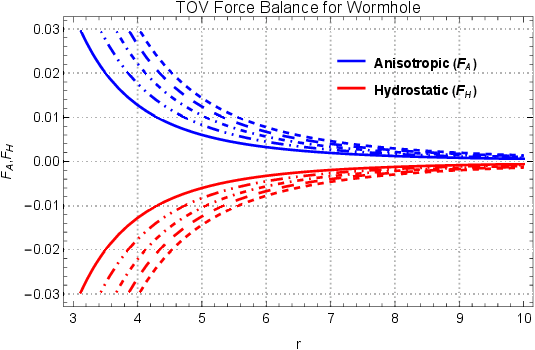} \hspace{1em}\includegraphics[width=6.5cm,height=5cm]{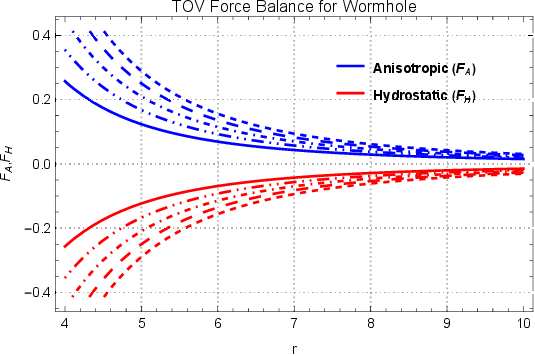}\\
    \includegraphics[width=6.5cm,height=5cm]{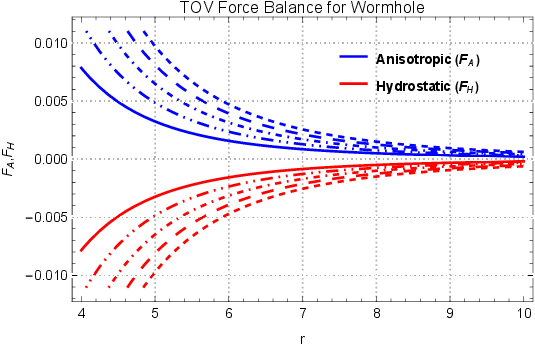} \hspace{1em}
    \caption{Here, two different forces ($F_A$ and $F_H$)  on the wormhole spacetimes are shown against the radial coordinate $r$ for all three cases [\eqref{e24},\eqref{e29},\eqref{e39}]. The values of the parameters used here are the same for drawing the profiles.} \label{fig17}   
\end{figure*}

To study the stability of the wormhole, we plot two forces $F_H$ and $F_A$ against the radial coordinate $r$ and inspect their evolution in fig. [\ref{fig17}]. It is seen that a stable configuration matches a specific dissolution of these forces, i.e., the total force becomes extinct all over the wormhole geometry. We see that the anisotropic force $F_A$ introduces positive behaviour, whereas the hydrostatic force $F_H$ introduces negative behaviour, i.e., both forces are identical but opposite to each other in sign. Fig.[\ref{fig17}] illustrates that the forces counterbalance each other, maintaining the system in equilibrium and resulting in stable wormholes.

\section{Conclusion}
The introduction of dark energy addresses the problem of the universe's rapid expansion, which may be equivalently addressed by considering modifications of General Relativity. Modified theories of gravity can effectively address the dark-matter issue as well. Fundamentally, there are two different approaches to altering General Relativity (GR): either the matter content or the geometrical properties of the action of this theory can be changed. This study examines a particular modification of general relativity as $f(Q)$ gravity, which is known as symmetric teleparallel gravity. Since $f(Q)$ gravity is a highly altered theory that has recently been constructed, much research is being done to examine its present applications in cosmology. Furthermore, the analysis of Wormhole solutions in $f(Q)$ gravity is a relatively new study. It is known that the NEC needs to be violated to have a traversable WH. Although physically impractical, exotic matter can be present in the WH-throat and make the NEC violation possible. In this work, three WH solutions were investigated for a linear form of $f(Q)$ theory. Furthermore, to make our computations easier, we considered the function $\phi(r)$ equal to zero.

The two forms for the varying Chaplygin gas that we examined in the first section of the study are $P_r=\frac{B b(r)^u}{\rho^\alpha}$ and $P_r=-\frac{B}{\rho^\alpha}$, respectively. For the changing barotropic fluid, we expressed the radial pressure as $p_r=-\omega \rho b(r)^u$ where $b(r)$ is the shape function. We were able to acquire accurate wormhole solutions in all of the scenarios that were examined. The requirements are met by the shape function as demonstrated through specific examples. Furthermore, the analysis of the first wormhole model, which is represented by the formula $P_r=\frac{B b(r)^u}{\rho^\alpha}$ for the radial pressure, have revealed that it is possible to violate the NEC, or ($\rho+P_r$). We computed the power-law form of the shape function $b(r)$. To meet the asymptotically flatness criterion, the parameter $\alpha$ must be negative. In light of this, we have examined every prerequisite for a form function as well as energy conditions. Null energy condition is violated as the model varies with the energy density $\rho \geq 0$ and the radial pressure is negative all over the space.

Based on the analysis of $P_r = -\frac{B}{\rho^\alpha}$, the radial equation of state parameter exhibiting model behavior is expected to result in wormhole formation. One may anticipate that wormhole models are validated only in terms of the NEC through the $P_t$ and radial pressures $P_r$ at the throat. The second wormhole model with $P_r = -\frac{B}{\rho^\alpha}$ is analyzed and it is seen that the DEC and NEC, expressed in terms of $P_t$ and $P_r$ respectively, are being violated. All of the aforementioned energy criteria are fulfilled for the formation of the wormhole.

A similar image of the models defined by different barotropic fluids has been obtained from the investigation reported in the second half of the paper. Furthermore, we would like to point out that energy conditions such as $\rho\geq 0$ and radial pressure are satisfied everywhere,, and the NEC (Null Energy Condition) in terms of $P_r$ is violated everywhere. Studying a specific wormhole model, represented by $p_r = -\omega \rho b(r)^u$, yields the same quantitative research behavior. In the third model, we generate a special shape function for $b(r)$ using the EOS parameter.For all three models, we studied the stability of the system using the Tolman-Oppenheimer-Volkoff equation (TOV), and under different energy situations, we see that the NEC was violated. Near the wormhole's throat, exotic matter may be present, resulting in each model's violation of NEC. Therefore, it is logical to state that the appropriate geometries for wormholes in $f(Q)$ gravity were breached at their throats, exhibiting a violation of the NEC. In the future, in a more generalised $f(Q)$ format, wormhole geometries could be investigated, since this could aid in the construction of a wormhole without the need for specific conditions. However, we anticipate reporting on wormhole models made with more unusual variations of the Chaplygin gas in future projects. Numerous research studies support the idea that this dark-energy depiction has some use. It would thus seem sensible to investigate the development of wormholes with this matter composition as well.

\section*{Acknowledgments}
TC and PR acknowledge the hospitality and the research facilities of the Inter-University Centre for Astronomy and Astrophysics (IUCAA), Pune, India, during their scientific visit. PR acknowledges IUCAA for granting a visiting associateship. The authors thank the anonymous referee for his/her invaluable comments that helped them to improve the quality of the manuscript.

\section*{Data Availability Statement}

No data was generated or analyzed in this study.

\section*{Conflict of Interest}

There are no conflicts of interest.

\section*{Funding Statement}

There is no funding to report for this article.


\end{document}